\documentclass[aps,prd,nofootinbib,twocolumn,groupedaddress,floatfix]{revtex4}
\usepackage{graphicx}
\usepackage{epsfig}
\usepackage{dcolumn}
\usepackage{amsmath}
\def\Journal#1#2#3#4{{#1} {\bf#2}, #3 (#4)}

\def\NPA{{\rm Nucl. Phys.} A}

\def\PLB{{\rm Phys. Lett.}  B}
\def\PRL{\rm Phys. Rev. Lett.}
\def\PRD{{\rm Phys. Rev.} D}
\def\PRC{{\rm Phys. Rev.} C}

\def\ep{\epsilon}
\def\vep{\varepsilon}
\def\la{\langle}
\def\ra{\rangle}

\def\al{\alpha}

\def\be{\begin{equation}}
\def\ee{\end{equation}}
\def\bea{\begin{eqnarray}}
\def\eea{\end{eqnarray}}
\input{epsfig.sty}
\begin{document}
\title{The Light-Front Zero-Mode Contribution to the Good Current 
in Weak Transitions}
\author{ Ho-Meoyng Choi$^{a}$ and Chueng-Ryong Ji$^{b}$\\
$^a$ Department of Physics Education, Kyungpook National University,
     Daegu, Korea 702-701\\
$^b$ Department of Physics, North Carolina State University,
Raleigh, NC 27695-8202}

\begin{abstract}
We examine the light-front zero-mode contribution to the good(+)
current matrix elements between pseudoscalar and vector mesons.
In particular, we discuss the transition form factor $f(q^2)$ which has 
been suspected to have the light-front zero-mode contribution. 
While the zero-mode contribution in principle depends on the form of the 
vector meson vertex $\Gamma^\mu=\gamma^\mu - (P_V -2k)^\mu/D$, 
the form factor $f(q^2)$ is found to be free from the 
zero-mode contribution if the denominator $D$ contains the term
proportional to the light-front energy $(k^-)^{n}$ 
with the power $n>0$.
The lack of zero-mode contribution benefits the light-front 
quark model phenomenology. We present our numerical calculations 
for the $B\to\rho$ transition.

\end{abstract}


\maketitle

\section{Introduction}
{\label{sect.I}}
For its simplicity and predictive power, the light-front
constituent quark model(LFQM) appears to be a useful phenomenological tool 
to study various electroweak properties of 
mesons~\cite{Te, Dz,CCP,CC,Ja90,Mel96,Mel01,HY97,CJ99,CJ97,MF97}.
The simplicity of the light-front(LF) quantization~\cite{BPP} is 
essentially attributed to the suppression of the vacuum fluctuations with 
the decoupling 
of complicated zero-modes and the conversion of the dynamical problem from 
boost to rotation. The suppression of vacuum fluctuations is due to the 
rational energy-momentum dispersion relation which correlates the signs of 
the LF energy $k^-=k^0-k^3$ and the LF momentum $k^+= k^0+k^3$.

However, the zero-mode($k^+=0$) complication in the matrix 
element has been noticed for the electroweak form factors involving 
a spin-1 particle\cite{Ja99,Ja03,BCJ02,BCJ03,CJ04}.  
A growing concern~\cite{Ja99,Ja03,BCJ02,BCJ03,CJ04,CCH} 
is to pin down which form factors get the zero-mode contributions.
The zero-mode contributions can be interpreted as residues
of virtual pair creation processes in the 
$q^+(=q^0+q^3)\to 0$ limit~\cite{MFS}.
In the absence of zero-mode contributions, the hadron form factors can be 
obtained rather straightforwardly by just taking into account only the 
valence contributions (or the diagonal matrix elements in the 
LF Fock-state expansion). Thus, it is quite significant to resolve 
the issue related to the zero-mode contribution to the hadron form 
factors. 

In an effort to clarify this issue, Jaus~\cite{Ja99,Ja03} and
we~\cite{BCJ02,BCJ03,CJ04} independently investigated the spin-1 
electroweak form factors in the past few years.
Jaus~\cite{Ja99,Ja03} proposed a covariant LF approach
involving the lightlike four vector $\omega^\mu (\omega^2 = 0)$ as 
a variable and developed a way of finding the zero-mode contribution
to remove the spurious amplitudes proportional to $\omega^\mu$.
Our formulation, however, is intrinsically distinguished from this
$\omega$-dependent formulation since it involves neither $\omega^\mu$ 
nor any unphysical form factors. Our method of finding the zero-mode 
contribution is a direct power-counting of the longitudinal momentum
fraction in $q^+\to 0$ limit for the off-diagonal elements in the 
Fock-state expansion of the current matrix\cite{BCJ02,BCJ03,CJ04}. 
Since the longitudinal momentum fraction is one of the integration 
variables in the LF matrix elements ({\it i.e.} helicity amplitudes),
our power-counting method is straightforward as far as we know
the behaviors of the longitudinal momentum fraction in the integrand.  
When the manifestly covariant model for the 
vector meson vertex $\Gamma^\mu$ is available, we have confirmed that 
the results found our way coincide with the ones from 
the manifestly covariant calculation.

\begin{figure}
\includegraphics[width=1.5in]{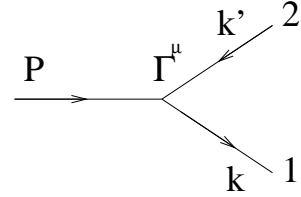}
\caption{Diagramatic representation of the vector meson coupling
$V-q\bar{q}$.}
\label{fig1}
\end{figure}

For a rather simple (manifestly covariant) vertex $\Gamma^\mu
=\gamma^\mu$, both Jaus and we agree on the absence of zero-mode 
contributions to the spin-1 electroweak form factors. 
However, Jaus and we do not 
agree when $\Gamma^\mu$ is extended to the more phenomenologically
accessible ones given by
\be\label{eq1}
\Gamma^\mu=\gamma^\mu
-\frac{(k+k')^\mu}{D},
\ee
where $k$ and $k'$ are the relative four momenta
for the constituent quark 1 and anti-quark 2 as shown in Fig.~\ref{fig1}.
Although Jaus's calculation and our calculation used the same 
denominator $D$ in Eq.(\ref{eq1}), 
they led to the different conclusions in the analysis of the zero-mode 
contribution.
Even if $D$ is chosen in such a way to get the manifestly covariant
$\Gamma^\mu$, the difference in the conclusions doesn't go away.

For the spin-1 elastic form factor calculations, 
Jaus's conclusion\cite{Ja99,Ja03} is that the matrix elements
$\la h'=0|J^+|h=1 \ra$ and $\la h'=0|J^+|h=0\ra$ both get the zero-mode 
contributions so that one cannot avoid
the zero-mode contributions to the form factor $F_2(q^2)$ 
for the vector meson.   
However, we recently~\cite{CJ04} found that only the matrix element
$\la h'=0|J^+|h=0\ra$ gets the zero-mode contribution so that 
we can avoid the zero-mode contribution to $F_2(q^2)$
without using the matrix element $\la h'=0|J^+|h=0\ra$.
While this calls for a clarification whether the $\omega$-dependent
formulation adds the more complication in the effect of zero-modes,
our finding of zero-mode contribution only in $\la h'=0|J^+|h=0\ra$
is quite significant in the LFQM phenomenology. It opens up
a possibility to make reliable predictions on the spin-1 
elastic form factors as we presented in the example
of the $\rho$ meson\cite{CJ04}.

Similarly, for the weak transition form factors between the 
pseudoscalar(P) and vector(V) mesons, Jaus\cite{Ja99,Ja03} concluded 
that the form factor $A_1(q^2)$[or $f(q^2)$] receives the zero-mode
contribution\footnote{We are not concerned with the form factor
$a_-(q^2)$ since the zero-mode contribution to the bad(-) 
current is not unexpected. Here, we discuss the zero-mode contribution 
to the good(+) current matrix elements only.}. 
Our aim of this work is to examine the zero-mode issue 
of this form factor $f(q^2)$ using our method. As we show in this work,
we again do not agree with his result but find that $f(q^2)$ is free
from the zero-mode contribution if the denominator $D$ in 
Eq.(\ref{eq1}) contains the term proportional to the LF energy 
$(k^-)^{n}$ with the power $n>0$. 
The phenomenologically accessible LFQM satisfies this condition
$n>0$.

In this work, we shall compute the weak transition form factors between 
pseudoscalar and vector mesons in three typical cases  
of the vector meson vertex, {\it i.e.}\\  
(1) $D=D_{\rm cov}(M_V)\equiv M_V + m_q+m_{\bar q}$,
where $M_V$ is the physical vector meson mass;\\ 
(2) $D=D_{\rm cov}(k\cdot P)\equiv [2 k\cdot P + M_V(m_q+ m_{\bar q}) 
-i\epsilon]/M_V$,
where $P$ is four momentum of the vector meson\cite{MF97};\\
(3) $D=D_{LF}(M_0)\equiv M_0 + m_q + m_{\bar q}$~\cite{Ja90},
where $m_q(m_{\bar q})$ is the mass for the consituent quark(anti-quark) 
and $M_0$ is the invariant mass of the vector meson.

For the manifestly covariant cases (1) and (2), we shall analyze two
different LF frames ($q^+=0$ and $q^+\neq 0$) and confirm the 
frame-independence of the physical observables. In the case (3), however, 
$D_{LF}(M_0)$ does not yield a manifestly covariant
$\Gamma^\mu$ and thus we cannot compute the nonvalence contribution
involving the non-wavefunction vertex beyond the two-body Fock state. The
non-wavefunction vertex which satisfies the requirement that the physical
observables must be Lorentz invariant has not yet been realized in the
case (3). Nevertheless, the lessons from the manifestly covariant cases
({\it e.g.} (1) and (2)) provide a significant constraint on the
non-wavefunction vertex, namely the power $n$ of the LF energy $(k^-)^n$
should be common both in the valence and nonvalence contributions. We
don't see any reason why this constraint cannot be applied to the case
(3). The continuity of the power $n$ between the valence 
and nonvalence contributions is sufficient for us to show the absence of the 
zero-mode contribution using the power counting of the longitudinal momentum
fraction in $q^+\to 0$ limit. The absence of zero-mode contribution 
assures that our valence result in 
$q^+ = 0$ frame is the full result in the case (3).  

The paper is organized as follows. In Sec.II, we present the 
Lorentz-invariant weak form factors between pseudoscalar and vector mesons
and the kinematics for the reference frames used in our analysis.
We also briefly discuss Jaus's approach.  
In Sec.III, we present our LF calculation of the weak transtion form factors
and discuss the criterion for the existence/nonexistence of the zero-mode.
In Sec.IV, we present our numerical results for the weak transition form 
factors in the above three cases;(1)$D_{\rm cov}(M_V) (n=0)$,
(2)$D_{\rm cov}(k\cdot P) (n=1)$,(3)$D_{LF}(M_0) (n=1/2)$. 
We compare them with the results obtained from Jaus's method. Conclusions 
follow in Sec.V. In the Appendix, the trace term to compute the 
nonvalence contribution is summarized.

\section{Model Description}
The Lorentz-invariant transition form factors\footnote{The transition 
form factors defined in Eq.~(\ref{eq:1}) are often
given by the following convention~\cite{BSW},
\begin{eqnarray*}
V(q^2) &=& (M_P + M_V)g(q^2), \nonumber \\
A_1(q^2) &=& {f(q^2) \over {M_P + M_V}}, \nonumber \\
A_2(q^2) &=& -(M_P + M_V)a_{+}(q^2), \nonumber \\
A_0(q^2) &=& \frac{1}{2 M_V} \biggl[f(q^2)+ (M_P^2-M_V^2)a_{+}(q^2)
+ q^2 a_{-}(q^2) \biggr],
\nonumber\\
{\label{eq:2}}
\end{eqnarray*}
where $M_P$ and $M_V$ are the physical pseudoscalar and vector meson 
masses, respectively.}  
$g$, $f$, $a_{+}$, and
$a_{-}$ between a pseudoscalar meson with four-momentum $P_1$
and a vector meson with four-momentum $P_2$ and helicity $h$ are
defined~\cite{AW} by the matrix elements of the electroweak current
$J^\mu_{V-A} = V^\mu - A^\mu$ from the
initial state $|P_1;00\ra$ to the final state $|P_2;1h\ra$:
\begin{eqnarray}
&&\la P_2;1h|J^\mu_{V-A}|P_1;00\ra
\nonumber\\
&&\hspace{0.5cm} = 
i g(q^2) \varepsilon^{\mu\nu\alpha\beta} 
\epsilon^*_{\nu}P_\alpha q_\beta -f(q^2) \epsilon^{*\mu}
\nonumber\\
&&\hspace{0.5cm} -a_{+}(q^2)(\epsilon^{*}\cdot P)
P^\mu -a_{-}(q^2)(\epsilon^{*}\cdot P)q^\mu,
{\label{eq:1}}
\end{eqnarray}
where the sum of $P_1^\mu$ and $P_2^\mu$ is denoted by $P^\mu$, the 
momentum transfer $q^\mu$ is given by $q^\mu = P_1^\mu - P_2^\mu$, 
and the polarization vector $\epsilon^*=\epsilon^*(P_2,h)$ of the
final state vector meson satisfies the Lorentz condition $\epsilon^* 
(P_2,h) \cdot P_2 = 0$.
While the form factor $g(q^2)$ is associated with the vector
current $V^\mu$, the rest of 
the form factors 
$f(q^2)$, $a_{+}(q^2)$, and
$a_{-}(q^2)$ are coming from the axial-vector current $A^\mu$.
The polarization vectors used in this analysis are given by
\vspace{-0.2cm}
\bea{\label{pol_vec}}
\epsilon^\mu(\pm 1)&=&[\ep^+,\ep^-,\ep_\perp]
=\biggl[0,\frac{2}{P^+_2}{\bf\epsilon}_\perp(\pm)\cdot{\bf P_{2\perp}},
{\bf\epsilon}_\perp(\pm 1)\biggr],
\nonumber\\
{\bf\epsilon}_\perp(\pm 1)&=&\mp\frac{(1,\pm i)}{\sqrt{2}},
\nonumber\\
\epsilon^\mu(0)&=&
\frac{1}{M_2}\biggl[P^+_2,\frac{{\bf P}^2_{2\perp}-M^2_2}{P^+_2},
{\bf P}_{2\perp}\biggr].
\eea
\vspace{-0.1cm}
\begin{figure}
\includegraphics[width=3.0in]{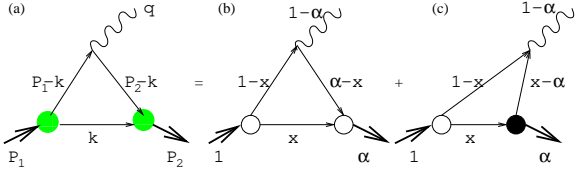}
\caption{The covariant diagram (a) corresponds to the sum of the
LF valence diagram (b) and the nonvalence diagram (c). The large white
and black blobs at the meson-quark vertices in (b) and (c) represent
the ordinary LF wave function and the non-wave function vertex,
respectively.}
\label{fig2}
\end{figure}

The covariant diagram in Fig.~\ref{fig2}(a) for the transition form factors 
between pseudoscalar and vector mesons is in general
equivalent to the sum of LF valence diagram (b) and the
nonvalence diagram (c), where $\al=P_2^+/P_1^+=1-q^+/P_1^+$.
From the covariant diagram of Fig.~\ref{fig2}(a), the matrix element 
$\la J^\mu_{V-A}\ra_h\equiv\la P_2;1h|J^\mu_{V-A}|P_1;00\ra$ is given by
\begin{equation}
\la J^\mu_{V-A}\ra_h 
= ig_1g_2\int\frac{d^4k}{(2\pi)^4}
\frac{S_{\Lambda_1}(P_1-k)S^\mu_h S_{\Lambda_2}(P_2-k)}
{S_{m_1} S_m S_{m_2}},
{\label{eq:4}}
\end{equation}
where $g_1$ and $g_2$ are the normalization factors which can be fixed by
requiring charge form factors of pseudoscalar and vector mesons to be
unity at $q^2=0$, respectively. Following the previous work~\cite{BCJ1},
we replaced the point gauge-boson vertex $\gamma^\mu (1-\gamma_5)$ by a 
non-local(smeared) gauge-boson vertex
$S_{\Lambda_1}(P_1-k)\gamma^\mu (1-\gamma_5)
S_{\Lambda_2}(P_2-k)$ to regularize the covariant fermion triangle loop
in ($3+1$) dimensions, where $S_{\Lambda_i}(P_i)
=\Lambda^2_i/(P_i^2-{\Lambda_i}^2+i\vep)$ and 
$\Lambda_i$ plays the role of momentum cut-off similar
to the Pauli-Villars regularization.  The rest of the
denominators in Eq.~(\ref{eq:4}) coming from the intermediate fermion 
propagators in the triangle loop diagram are given by
\begin{eqnarray}
S_{m_1} &=& p_1^2 -{m_1}^2 + i\vep, \nonumber \\
S_m &=& k^2 - m^2 + i\vep, \nonumber \\
S_{m_2} &=& p_2^2 -{m_2}^2 + i\vep,
{\label{eq:5}}
\end{eqnarray}
where $m_1$, $m$, and  $m_2$ are the masses of the constituents
carrying the intermediate four-momenta $p_1=P_1 -k$, $k$, and $p_2=P_2
-k$, respectively. 

The trace term in Eq.~(\ref{eq:4}), $S^\mu_h$, is given by
\begin{equation}
S^\mu_h = {\rm Tr}[(\not\!p_2 + m_2)\gamma^\mu (1-\gamma_5)
(\not\!p_1 +m_1)\gamma_5(-\not\!k + m)\epsilon^*\cdot\Gamma],
{\label{eqq:4}}
\end{equation}
where the final state vector meson vertex operator $\Gamma^\mu$ 
is given by
\begin{equation}
\Gamma^\mu=\gamma^\mu-\frac{(P_2-2k)^\mu}{D}.
\label{eq5}
\end{equation}
We shall analyze the three different cases of $D$-term,{\it 
i.e.}
\bea\label{D-term}
&&(1) D_{\rm cov}(M_V)=M_V + m_2 + m,
\nonumber\\
&&(2) D_{\rm cov}(k\cdot P_2)
=\frac{[2k\cdot P_2 + M_V(m_2 + m) -i\epsilon]}{M_V},
\nonumber\\
&&(3) D_{LF}(M'_0)=M'_0 + m_2 + m, 
\eea
where the prime denotes the final state.

In our trace term calculation, we separate Eq.~(\ref{eqq:4}) into the
on-mass-shell propagating part $S^\mu_{\rm on}$ and the off-mass-shell 
part $S^\mu_{\rm off}$,{\it i.e.}
\bea{\label{sep}}
S^\mu_h&=& (S^\mu_h)_{\rm on} + (S^\mu_h)_{\rm off},
\eea
via 
\bea{\label{ID}}
\not\!p + m &=&
(\not\!p_{\rm on} + m) + \frac{1}{2}\gamma^+(p^- - p^-_{\rm on}).
\eea
While the on-mass-shell part $(S^\mu_h)_{\rm on}$ indicates that all
three quarks are on their respective mass-shell,{\it i.e.} $k^-=k^-_{\rm 
on}$ and $p^-_{i}=p^-_{i\rm on}(i=1,2)$, the off-mass-shell part 
$(S^\mu_h)_{\rm off}$ includes the term proportional 
to $(k^- - k^-_{\rm on})$~\cite{BCJ03}. 
The trace terms $(S^+_h)_V$ and $(S^+_h)_A$ in Eq.~(\ref{eqq:4})
for the vector and axial-vector currents are given by
\begin{widetext}
\bea\label{S_VA}
(S^{+}_h)_V &=& 4i\varepsilon^{+\mu\nu\al}
\biggl[
[ m_1(p_{2\rm on})_\mu (k_{\rm on})_\nu
- m_2 (p_{1\rm on})_\mu (k_{\rm on})_\nu
- m (p_{1\rm on})_\mu (p_{2\rm on})_\nu ]\epsilon^*_\al
-\frac{(p_2-k)\cdot\epsilon^*(h)}{D}
(p_{1\rm on})_\mu(p_{2\rm on})_\nu(k_{\rm on})_\al
\biggr]
\nonumber\\
(S^+_h)_A &=& 4 m_1[ (k_{\rm on}\cdot\epsilon^*)p^+_{2\rm on}
+ (p_{2\rm on}\cdot\epsilon^*)k^+_{\rm on}
- (p_{2\rm on}\cdot k_{\rm on})\epsilon^{*+} ]
- 4 m_2[ (k_{\rm on}\cdot\epsilon^*)p^+_{1\rm on}
- (p_{1\rm on}\cdot\epsilon^*)k^+_{\rm on}
+ (p_{1\rm on}\cdot k_{\rm on})\epsilon^{*+} ]
\nonumber\\
&&+ 4 m[ (p_{2\rm on}\cdot\epsilon^*)p^+_{1\rm on}
+ (p_{1\rm on}\cdot\epsilon^*)p^+_{2\rm on}
- (p_{1\rm on}\cdot p_{2\rm on})\epsilon^{*+} ]
- 4 m_1 m_2 m \epsilon^{*+}
-4(k^- - k^-_{\rm on})m_2p^+_{1\rm on}\epsilon^{*+}
\nonumber\\
&&+4\frac{(p_2-k)\cdot\epsilon^*(h)}{D}
\biggl[ (p_{2\rm on}\cdot k_{\rm on}-m_2m)p^+_1
+(p_{1\rm on}\cdot k_{\rm on}+m_1m)p^+_2
- (p_{1\rm on}\cdot p_{2\rm on}+m_1m_2)k^+
\nonumber\\
&&\hspace{3cm}+ (k^--k^-_{\rm on})p^+_{1\rm on}p^+_{2\rm on}\biggr].
\eea
\end{widetext}

A different approach calculating Eq.~(\ref{eqq:4}) can be found in 
Refs~\cite{Ja99,Ja03} where Jaus used the four-vector and tensor 
decompositions of the internal four-momentum $p_1$ including the 
lightlike four-vector $\omega$ in the trace terms,{\it e.g.} 
four-vector decomposition of $p_1$
is given by
$p_{1\mu}=A^{(1)}_1P_\mu + A^{(1)}_2q_\mu + C^{(1)}_1\omega_\mu$,
where $C$-type functions are $\omega$-dependent while $A$-type functions 
are $\omega$-independent. His main idea for the calculation of the
trace term is to separate the term proportional to $N_2=k^2-m^2$ from
the rest of the terms in the trace. The $\omega$-dependent $C$-type(and 
also $B$-type arising from the tensor decomposition of $p_{1\mu}p_{1\nu}$) 
functions include this $N_2$-term, 
{\it e.g.} $C^{(1)}_1=-N_2 + Z_2(x,{\bf k}_\perp)$. 
Although this $N_2$-term 
vanishes for the spectator quark with 
the momentum $k$ being on-mass-shell($k^-=k^-_{\rm on}$), it may give 
nonvanishing contribution if the spectator quark is off-mass-shell, 
{\it i.e.} $k^-=p^-_{1\rm on}$.
If this happens, then the nonvanishing $N_2$-term contribution related 
to the zero-mode contribution should be included to obtain the Lorentz 
invariant form factor.  Jaus discussed that the inclusion of the 
zero-mode(without involving higher Fock states or the nonvalence 
contributions) can be made by the replacement $N_2 \to Z_2$,{\it i.e.} 
$C^{(1)}\doteq 0$.
However, his $C^{(1)}\doteq 0$ prescription is valid
only at the particular choice of the vector meson vertex operator 
$\Gamma^\mu$ in Eq.~(\ref{eq5}),{\it e.g.} $C^{(1)}\doteq 0$ is valid only 
for the $D_{\rm cov}(M_V)$ in Eq.~(\ref{D-term}) but not for 
$D_{\rm cov}(k\cdot P_2)$ and $D_{LF}(M'_0)$ as we shall show in the 
following sections.  For the comparison with Jaus's $N_2$-term 
prescription later on, we note that his $N_2$-term corresponds to our 
$(k^--k^-_{\rm on})$-term via 
$N_2=k^2-m^2=k^+(k^- -k^-_{\rm on}) + k^2_{\rm on}-m^2=k^+(k^- -k^-_{\rm on})$.

\section{Light-front calculation of the weak form factors}

In the LF calculation of the weak form factors, we use  
${\bf P}_{1\perp}=0$ frame with the
(timelike) momentum transfer $q^2=(P_1-P_2)^2$ given by
\begin{equation}
{\label{tq2}}
q^2=q^+q^- - {\bf q}^2_\perp =(1-\al)\biggl(M^2_1 -
\frac{M^2_2}{\al}\biggr) -\frac{{\bf q}^2_\perp}{\alpha}.
\end{equation}
We shall use only the plus component of the $V-A$ current for the 
calculations of
LF valence[Fig.~\ref{fig2}(b)] and nonvalence [Fig.~\ref{fig2}(c)] diagrams.

\subsection{ Matrix elements of the weak current}
In the valence region $0<k^+<P^+_2$, the pole $k^-=k^-_{\rm on}=(m^2 +
{\bf k}^2_\perp -i\vep)/k^+$ ({\it i.e.}, the spectator quark) is located in
the lower half of the complex $k^-$-plane.  

Thus, the Cauchy formula for the $k^-$-integration in 
Eq.~(\ref{eq:4}) gives
\bea{\label{jval}}
\la J^{+}_{V-A}\ra^h_{\rm val}&=&
\frac{g_1g_2\Lambda^2_1\Lambda^2_2}{2(2\pi)^3}\int^\al_0
\frac{dx}{x}\int d^2{\bf k}_\perp
\nonumber\\
&&\times
\psi_i(x,{\bf k}_\perp)(S^{+}_h)_{\rm on}
\psi_f(x',{\bf k'}_\perp),
\eea
where
\bea\label{psi_nv}
\psi_i(x,{\bf k}_\perp)&=&
\frac{1}{(1-x)^2(M^2_1-M^2_0)(M^2_1-M^2_{\Lambda_1})},
\nonumber\\
\psi_f(x',{\bf k'}_\perp)&=&
\frac{1}{(1-x')^2(M^2_2-M'^2_0)(M^2_2-M'^2_{\Lambda_2})},
\nonumber\\
\eea
and
\bea
M^2_0&=& \frac{{\bf k}^2_\perp + m^2_1}{1-x}
+\frac{{\bf k}^2_\perp + m^2}{x},
\nonumber\\
M'^2_0&=& \frac{{\bf k'}^2_\perp + m^2_2}{1-x'}
+\frac{{\bf k'}^2_\perp + m^2}{x'},
\nonumber\\
M^2_{\Lambda_1}&=& M^2_0(m_1\to\Lambda_1),\;
M'^2_{\Lambda_2}= M'^2_0(m_2\to\Lambda_2).
\eea
The final state momentum variables are given by 
${\bf k'}_\perp={\bf k}_\perp + x'{\bf q}_\perp$ and $x'=x/\alpha$.
Note that the trace term in Eq.~(\ref{jval}) includes only the 
on-mass-shell propagating part since the pole structure 
$k^-=k^-_{\rm on}$(or equivalently $N_2=0$)
leads to the vanishing off-mass-shell contributions.

In the nonvalence region $P^+_2<k^+<P^+_1$, the poles at
$k^- = k^-_{m_1} \equiv
P^-_1 + [m^2_1 +({\bf k}_\perp -{\bf P}_{1\perp})^2 -i\vep]/(k^+-P^+_1)$
(from the struck quark propagator) and
$k^- = k^-_{\Lambda_1} \equiv P^-_1 
+ [\Lambda^2_1+({\bf k}_\perp -{\bf P}_{1\perp})^2 -i\vep]/(k^+-P^+_1)$
(from the smeared quark-photon vertex),
are located in the upper half of the complex $k^-$-plane.

Thus, the Cauchy integration over $k^-$ in Eq.~(\ref{eq:4}) 
gives
\bea{\label{jnv}}
&&\la J^+_{V-A}\ra^h_{\rm nv}
\nonumber\\
&&= 
\frac{g_1g_2\Lambda^2_1\Lambda^2_2}
{2(2\pi)^3(\Lambda^2_1-m^2_1)}
\int^1_\al\frac{dx}{xx''(1-x'')(x-\al)}
\nonumber\\
&&\times
\int d^2{\bf k}_\perp
\biggl\{
\frac{S^+_{h}(k^-_{\Lambda_1})}
{(M^2_1-M^2_{\Lambda_1})(q^2-M^2_{\Lambda_1\Lambda_2})
(q^2-M^2_{\Lambda_1m_2})}
\nonumber\\
&&\;\;\;\;
- \frac{S^+_{h}(k^-_{m_1})}
{(M^2_1-M^2_{0})(q^2-M^2_{m_1\Lambda_2})(q^2-M^2_{m_1m_2}))}
\biggr\},
\eea
where
\bea
M^2_{\Lambda_1\Lambda_2}&=&
\frac{{\bf k''}^2_\perp+\Lambda^2_1}{x''}
+\frac{{\bf k''}^2_\perp + \Lambda^2_2}{1-x''},
\nonumber\\
M^2_{\Lambda_1 m_2}&=&
\frac{{\bf k''}^2_\perp+\Lambda^2_1}{x''}
+\frac{{\bf k''}^2_\perp + m^2_2}{1-x''},
\nonumber\\
M^2_{m_1 m_2}&=&
\frac{{\bf k''}^2_\perp+m^2_1}{x''}
+\frac{{\bf k''}^2_\perp + m^2_2}{1-x''},
\nonumber\\
M^2_{m_1\Lambda_2}&=&
\frac{{\bf k''}^2_\perp+m^2_1}{x''}
+\frac{{\bf k''}^2_\perp + \Lambda^2_2}{1-x''},
\eea
and 
\bea
x''&=& \frac{1-x}{1-\al}, \;
{\bf k''}_\perp = {\bf k}_\perp + x''{\bf q}_\perp.
\eea
The explicit forms of the trace terms $S^+_h(k^-_{m_1})$ and
$S^+_h(k^-_{\Lambda_1})$ are given in the Appendix.
In general, the trace terms in the nonvalence diagram include the
off-mass-shell contributions(or equivalently $N_2\neq 0$),{\it e.g.} 
$S^+_h(k^-_{m_1})=(S^+_h)_{\rm on} + (S^+_h)_{\rm off}(k^-_{m_1})$.

\subsection{ Extraction of weak form factors}
From Eqs.~(\ref{eq:1}),~(\ref{pol_vec}) and~(\ref{tq2}),
one obtains the relations between the current matrix elements and
the weak form factors as follows
\bea{\label{ch:4}}
\la J^+_V\ra^{h=1} &=&
-\frac{P^+_1}{\sqrt{2}}\varepsilon^{+-xy}q^L g(q^2),
\nonumber\\
\la J^+_V\ra^{h=0} &=& 0,
\eea
for the vector current and
\bea{\label{ch:5}}
\la J^+_A\ra^{h=1}
&=& \frac{P^+_1q^L}{\al\sqrt{2}}
\biggl[ (1+\al) a_+(q^2)
+ (1-\al) a_-(q^2)\biggr],
\nonumber\\
\la J^+_A\ra^{h=0} &=& \frac{\al P^+_1}{M_2}f(q^2)
+ \frac{\al P^+_1}{2M_2}
\biggl(M^2_1 - \frac{M^2_2}{\al^2}
+ \frac{{\bf q}^2_\perp}{\al^2}\biggr)
\nonumber\\
&\times&\biggl[ (1+\al)a_+(q^2) + (1-\al)a_-(q^2)\biggr],
\eea
for the axial-vector current. Here, $q^L=q_x - iq_y$. 

The extraction of weak form factors can be made in various ways. Among
them, there are two popular ways of extracting the form factors,
{\it e.g.} one can obtain the form factors (1) in the spacelike region 
using the $q^+ =0$ frame and then analytically continue to the timelike
region by changing ${\bf q}_\perp$ to $i{\bf q}_\perp$ in the form factor, 
or (2) in a direct timelike region using a $q^+>0$ frame. 

In this work, we shall use both the $q^+=0$ 
frame ($q^2=-{\bf q}^2_\perp$) and the purely longitudinal momentum 
frame ($q^+> 0$ and ${\bf q}_\perp=0$) where 
\bea{\label{pl:1}}
q^2&=&q^+q^- =(1-\al)\biggl(M^2_1 - \frac{M^2_2}{\al}\biggr).
\eea
\noindent
For this particular choice of the purely longitudinal momentum frame,
there are two solutions of $\al$ for a given $q^2$,{\it i.e.}
\begin{equation}{\label{pl:2}}
\al_{\pm}=\frac{M_2}{M_1}\biggl[
\frac{M^2_1 + M^2_2-q^2}{2M_1M_2} \pm
\sqrt{ \biggl(\frac{M^2_1 + M^2_2-q^2}{2M_1M_2}\biggr)^2-1}
\biggr],
\end{equation}
where the $+(-)$ sign in Eq.~(\ref{pl:2}) corresponds to the daugther
meson recoiling in the positive(negative) $z$-direction relative to
the parent meson. 
At the zero recoil ($q^2=q^2_{\rm max}$) and the maximum
recoil ($q^2=0$), $\alpha_\pm$ are respectively given by
\bea{\label{apm}}
\alpha_+(q^2_{\rm max})&=&\alpha_-(q^2_{\rm max})=\frac{M_2}{M_1},
\nonumber\\
\alpha_+(0)&=&1,\;\;
\alpha_-(0)=\biggl(\frac{M_2}{M_1}\biggr)^2.
\eea
The form factors should in principle be independent of the recoil directions
($\alpha_\pm$) if the nonvalence contributions are added to the valence
ones.

While the form factor $g(q^2)$ in the $q^+>0$ frame can be obtained
directly from Eq.~(\ref{ch:4}), the form factor $f(q^2)$ can be
obtained only after $a_\pm(q^2)$ are calculated.
To illustrate this, we define
\bea{\label{pl:3}}
\la J^+_A\ra^{h=1} |_{\al=\al_\pm}
&\equiv& \frac{P^+_1 q^L}{\sqrt{2}}I^+_A(\al_\pm).
\eea
Then we obtain from Eq.~(\ref{ch:5})
\bea{\label{pl:4}}
a_+(q^2)&=&\frac{
\al_+(1-\al_-)I^+_A(\al_+) - \al_-(1-\al_+) I^+_A(\al_-)}
{2(\al_+ -\al_-)},
\nonumber\\
a_-(q^2)&=&-\frac{
\al_+(1+\al_-)I^+_A(\al_+) - \al_-(1+\al_+) I^+_A(\al_-)}
{2(\al_+ -\al_-)},
\nonumber\\
\eea
and
\bea{\label{pl:5}}
f(q^2)&=& \frac{M_2}{\al P^+_1}\la J^+_A\ra^{h=0}
-\frac{1}{2}\biggl(M^2_1-\frac{M^2_2}{\al^2}\biggr)
\nonumber\\
&&\times\biggl[(1+\al)a_+(q^2) + (1-\al)a_-(q^2) \biggr].
\eea

As can be seen from Eqs. (\ref{ch:4}) and (\ref{ch:5}), 
one should be careful in setting ${\bf q}_\perp = 0$ to get the correct 
results in the purely longitudinal frame.  One cannot simply set 
${\bf q}_\perp=0$ from the start, but should set it to zero only after the 
form factors are extracted.

In the $q^+>0$ frame where ${\bf q}_\perp\neq 0$, 
the valence contribution to $g(q^2)$ is given by
\begin{widetext}
\bea\label{gval}
g_{\rm val}(q^2)&=&\frac{N}{8\pi^3}
\int^\al_0
\frac{dx}{x}\int d^2{\bf k}_\perp\;
\psi_i\psi_f
\biggl\{ {\cal A}_p +
\frac{{\bf k}_\perp\cdot{\bf q}_\perp}{{\bf q}^2_\perp}
[ \al(m_1-m) + (m-m_2) ]
+ \frac{2}{D}\biggl[ {\bf k}^2_\perp
-\frac{({\bf k}_\perp\cdot{\bf q}_\perp)^2}{{\bf q}^2_\perp}
\biggr]
\biggr\},
\eea
\end{widetext}
where $N=g_1g_2\Lambda^2_1\Lambda^2_2$ and ${\cal A}_p = xm_1 + (1-x)m$.
The form factor $g^{DY}(q^2)$ in $q^+=0$ (or Drell-Yan(DY)) frame is
given by
\begin{equation}
g^{DY}(q^2) = \lim_{\alpha\to 1}g_{\rm val}(q^2).
\label{gDY}
\end{equation}
Our result for $g^{DY}(q^2)$ is the same as the one obtained by 
Jaus(see Eq.~(4.13) in \cite{Ja99}).  Note that one needs to replace
$x$ by $(1-x)$ and ${\bf q}_\perp$ by $-{\bf q}_\perp$ between the  
two formulations to compare each other directly. The form factor $g(q^2)$ 
is found to be free from the zero-mode contribution.  The nonvalence 
contribution to 
$g(q^2)$ in $q^+>0$ frame can be obtained from Eq.~(\ref{jnv}) with the 
trace term given by Eq.~(\ref{A1}). 

The valence contribution to the matrix elements $I^+_A(\alpha)$ 
for $a_{\pm}(\alpha)$ in Eq.~(\ref{pl:4}) is given by
\begin{widetext}
\bea\label{Ia_v}
(I^+_A)_{\rm val}(\al)&=& \frac{2N}{8\pi^3}
\int^{\al}_0 \frac{dx}{x} \int d^2{\bf k}_\perp\; \psi_i \psi_f
\biggl\{
(1-2x'){\cal A}_p + \frac{{\bf k}_\perp\cdot{\bf q}_\perp}{{\bf q}^2_\perp}
[ (\al-2x)(m_1-m) - (m_2+m)]
\nonumber\\
&&\hspace{1cm}-\frac{2}{x'D}
\biggl(x' + \frac{{\bf k}_\perp\cdot{\bf q}_\perp}{{\bf q}^2_\perp}\biggr)
\biggl({\bf k}_\perp\cdot{\bf k'}_\perp
+ {\cal A}_p[(1-x')m - x'm_2]\biggr)
\biggr\}.
\eea
\end{widetext}
The form factor $a^{DY}_+(q^2)$ in $q^+=0$ frame is given by
\begin{equation}
a^{DY}_+(q^2)=\lim_{\al\to 1}\frac{(I^+_A)_{\rm val}(\al)}{2}.
\label{aDY}
\end{equation}
Our result of $a^{DY}_+(q^2)$ is the same as the one obtained 
by Jaus(see Eq.~(4.14) 
in~\cite{Ja99}). The form factor $a_+(q^2)$ 
is also found to be free from the zero-mode contribution.
The nonvalence contribution to $a_+(q^2)$ in $q^+>0$ frame can be obtained 
from Eq.~(\ref{jnv}) with the trace term given by Eq.~(\ref{A2}).
 
To obtain the form factor $f(q^2)$ in Eq.~(\ref{pl:5}), we need to compute
the matrix element involving the helicity zero,{\it i.e.}
$\la J^+_A\ra^{h=0}$. The explicit form of the valence contribution to
$\la J^+_A\ra^{h=0}$ is given by
\begin{widetext}
\bea\label{JAh0_val}
\la J^+_A\ra^{h=0}_{\rm val}&=&
-\frac{2N}{8\pi^3}\frac{P^+_1}{M_2}
\int^\alpha_0 \frac{dx}{xx'}\int \;d^2{\bf k}_\perp\; \psi_i\psi_f
\biggl\{ {\cal A}_p[x'(1-x')M^2_2 + m_2m + x'^2{\bf q}^2_\perp] 
+ {\bf k}^2_\perp(xm_1 + m_2 - xm)
\nonumber\\
&& + x'{\bf k}_\perp\cdot{\bf q}_\perp[2x(m_1-m)+m_2+m]
-\frac{(x'^2M^2_2-{\bf k'}^2_\perp-m^2)}{D}
\biggl( {\bf k}_\perp\cdot{\bf k'}_\perp 
+ {\cal A}_p [(1-x')m - x'm_2]\biggr)
\biggr\}.
\eea
\end{widetext}
The form factor $f^{DY}(q^2)$ in $q^+=0$ frame obtained from 
Eq.~(\ref{ch:5}),{\it i.e.}
\be{\label{fDY}} 
 f^{DY}(q^2)= -(M^2_1 - M^2_2 + {\bf q}^2_\perp)a^{DY}_+(q^2)
+\frac{M_2}{P^+_1}\lim_{\al\to 1}\la J^+_A\ra^{h=0}_{\rm val},
\ee
is explicitly given by
\begin{widetext}
\bea\label{fCJ}
f^{DY}(q^2)&=&-\frac{N}{8\pi^3}\int^1_0\frac{dx}{x^2}
\int\;d^2{\bf k}_\perp\; \psi_i\psi_f
\biggl\{
2{\bf k}^2_\perp(xm_1 + m_2 - xm)
+ 2x{\bf k}_\perp\cdot{\bf q}_\perp [ 2x(m_1 - m) + m_2 + m]
\nonumber\\
&&\; + {\cal A}_p [ x(1-2x)M^2_1 + xM^2_2 + 2m_2 m + x{\bf q}^2_\perp ]
+ x [q\cdot P+ {\bf q}^2_\perp][(1-2x)(m_1-m)-m_2-m]
\frac{{\bf k}_\perp\cdot{\bf q}_\perp}{{\bf q}^2_\perp}
\nonumber\\
&&\; -\frac{2}{xD}\biggl( {\bf k}_\perp\cdot{\bf k}'_\perp 
+ {\cal A}_p[(1-x)m - xm_2]\biggr)
\biggl(x [q\cdot P + {\bf q}^2_\perp]
[x+\frac{{\bf k}_\perp\cdot{\bf q}_\perp}{{\bf q}^2_\perp}]
 + x^2 M^2_2 - {\bf k}'^2_\perp - m^2 
\biggr)
\biggr\},
\eea
\end{widetext}
where $q\cdot P=M^2_1 - M^2_2$.
We should note that our result for 
$\lim_{\al\to 1}\la J^+_A\ra^{h=0}_{\rm val}$(i.e. $f^{DY}(q^2)$) is
different from the result obtained by the LF
formalism discussed in the Appendix C of Ref.\cite{Ja03}
(see,{\it e.g.} Eq.~(C2) in~\cite{Ja03}). That formalism\cite{Ja03} 
requires all quarks to be on their respective mass-shells and
replaces the physical vector meson mass $M_2$ in Eq.~(\ref{pol_vec}) by
the invariant meson mass $M'_0$. However, our result is obtained
by requiring only the struck quark($m_2$) to be on-mass-shell and using the
physical vector meson mass $M_2$ in Eq.~(\ref{pol_vec}).
The nonvalence contribution to $\la J^+_A\ra^{h=0}$ in $q^+>0$ frame
can be obtained from Eq.~(\ref{jnv}) with the trace term given by 
Eq.~(\ref{A3}) in our Appendix.

We now determine whether $f(q^2)$,{\it i.e.} $f^{DY}(q^2)$,
is free from the zero-mode.  The zero-mode contribution to $\la 
J^+_{A}\ra^h$ 
is defined as
\begin{equation}
\la J^+_{A}\ra^h_{\rm z.m.}
= \lim_{\alpha\to 1}\la J^+_{A}\ra^h_{\rm nv}.
\label{jzm1}
\end{equation}
To check if the zero-mode exists or not, we use the counting rule for
the factors of the longitudinal momentum fraction in 
Eq.~(\ref{jnv}),{\it e.g.} in the $q^+\to 0$ limit, the first term in 
Eq.~(\ref{jnv}) becomes
\bea{\label{jzm2}}
\la J^+_{A}\ra^h_{\rm z.m.}\hspace{-0.2cm} &\sim&\hspace{-0.2cm}
\lim_{\alpha\to 1}\int^1_\alpha dx
\frac{x(1-x)[x''(1-x'')]^2}{xx''(1-x'')(x-\alpha)}
S^+_h(k^-_{\Lambda_1})[\cdots ]
\nonumber\\
&=& \lim_{\alpha\to 1}\int^1_\alpha dx
\frac{(1-x)^2}{(1-\alpha)^2}
S^+_h(k^-_{\Lambda_1})[\cdots ]
\nonumber\\
&=& \lim_{\alpha\to 1}\int^1_0 dz
(1-\alpha)(1-z)^2 S^+_h(k^-_{\Lambda_1})[\cdots ],
\eea
where the variable change $x=\al + (1-\al)z$ was made and the terms in
$[\cdots]$ are regular in the $\al\to 1$(or equivalently
$x\to 1$) limit. The second term in Eq.~(\ref{jnv}) with
$S^+_h(k^-_{m_1})$ has the same power
counting of the longitudinal momentum fraction as the first term in
Eq.~(\ref{jzm2}). 

From Eq.~(\ref{jzm2}), one can determine the existence/non-existence of the 
zero-mode contribution to $f^{DY}(q^2)$ by counting the 
factors of the longitudinal momentum fraction, 
specifically $(1-x)$-factors, in the trace terms 
$S^+_{h=0}(k^-_{\Lambda_1})$ and $S^+_{h=0}(k^-_{m_1})$. Note that both
$k^-_{\Lambda_1}=k^-_{m_1}\sim p^-_{1\rm on}\sim 1/(1-x)$. Thus, from
$(S^+_{h=0})_A$ in Eq.~(\ref{S_VA}), the terms such as
$(k^- - k^-_{\rm on})p^+_{1\rm on}$ are regular for the 
factor $(1-x)$. All other on-mass-shell terms are also regular for the 
same factor $(1-x)$.  Thus, the only zero-mode suspected term is 
$(p_2-k)\cdot\epsilon^*(h=0)/D\sim 1/(1-x)D$. 

The power counting of $(1-x)$ in 
$(S^+_{h=0})(k^-_{\Lambda_1})$ depends on the vector meson 
vertices (See Eq.~(\ref{D-term})). We find that 
$(S^+_{h=0})(k^-_{\Lambda_1})$ is proportional to
(1) $(1-x)^{-1}=[(1-\al)(1-z)]^{-1}$ for $D_{\rm cov}(M_V)$,
(2) $(1-x)^0$ for $D_{\rm cov}(k\cdot P_2)$, 
and 
(3) $(1-x)^{-1/2}=[(1-\al)(1-z)]^{-1/2}$ for $D_{LF}(M'_0)$,
respectively.
These power-counting results show that the form factor $f^{DY}(q^2)$ 
receives the zero-mode contribution only for the $D_{\rm cov}(M_V)$
case but not for others. In fact, our power-counting should also hold in 
Jaus's case,{\it i.e.} $N_2/D$ for the zero-mode limit goes to (1) 
$Z_2/D$ for $D_{\rm cov}(M_V)$,
(2) $(1-x)N_2\doteq 0$ for $D_{\rm cov}(k\cdot P_2)$, 
and
(3) $\sqrt{(1-x)}N_2\doteq 0$ for $D_{LF}(M'_0)$, respectively. 
On the other hand, Jaus~\cite{Ja03,Ja99} used $N_2/D\doteq Z_2/D$ 
regardless of the $D$-terms and removed $C$-type functions as well as 
$B$-type functions. This explains how he reached the conclusion
that the form factor $f(q^2)$ receives the zero-mode contribution 
regardless of the vertices used in the model calculation.  
We have shown that his conclusion is correct for the case (1) 
(or $D_{\rm cov}(M_V)$) but not for the cases (2)(or $D_{\rm cov}(k\cdot 
P_2)$) and (3)(or $D_{LF}(M'_0)$). 
We now confirm our derivation through the numerical calculation in the 
next section.

\section{Numerical results}
In this section, we present the numerical results for the 
$B\to\rho$ transition form factors ($g(q^2),a_+(q^2),f(q^2)$) in the 
three different cases of meson vertex discussed above. We perform our LF 
calculation in the two different reference frames, {\it i.e.} $q^+=0$ and 
purely longitudinal $q^+>0$ frames.  We also compare our results with those 
obtained by Jaus~\cite{Ja99,Ja03}. 
We do not aim at finding the best-fit parameters to
describe the experimental data in this work. However, the
essential findings from the generic structure of our model 
calculations are expected to apply also for the
more realistic models, although the quantitative results would 
depend on the details of the model.
The model parameters for $B$ and $\rho$ mesons are taken same
as in Ref.\cite{BCJ03}:
$M_B = 5.28$ GeV, $M_\rho = 0.771$ GeV,
$m_b=4.9$ GeV, $\Lambda_b=10$ GeV, and $g_B=5.20$, 
as well as $m_u=m_d=0.43$ GeV, $\Lambda_{u(d)}=1.5$ GeV, and $g_\rho=5.13$.

\begin{figure*}
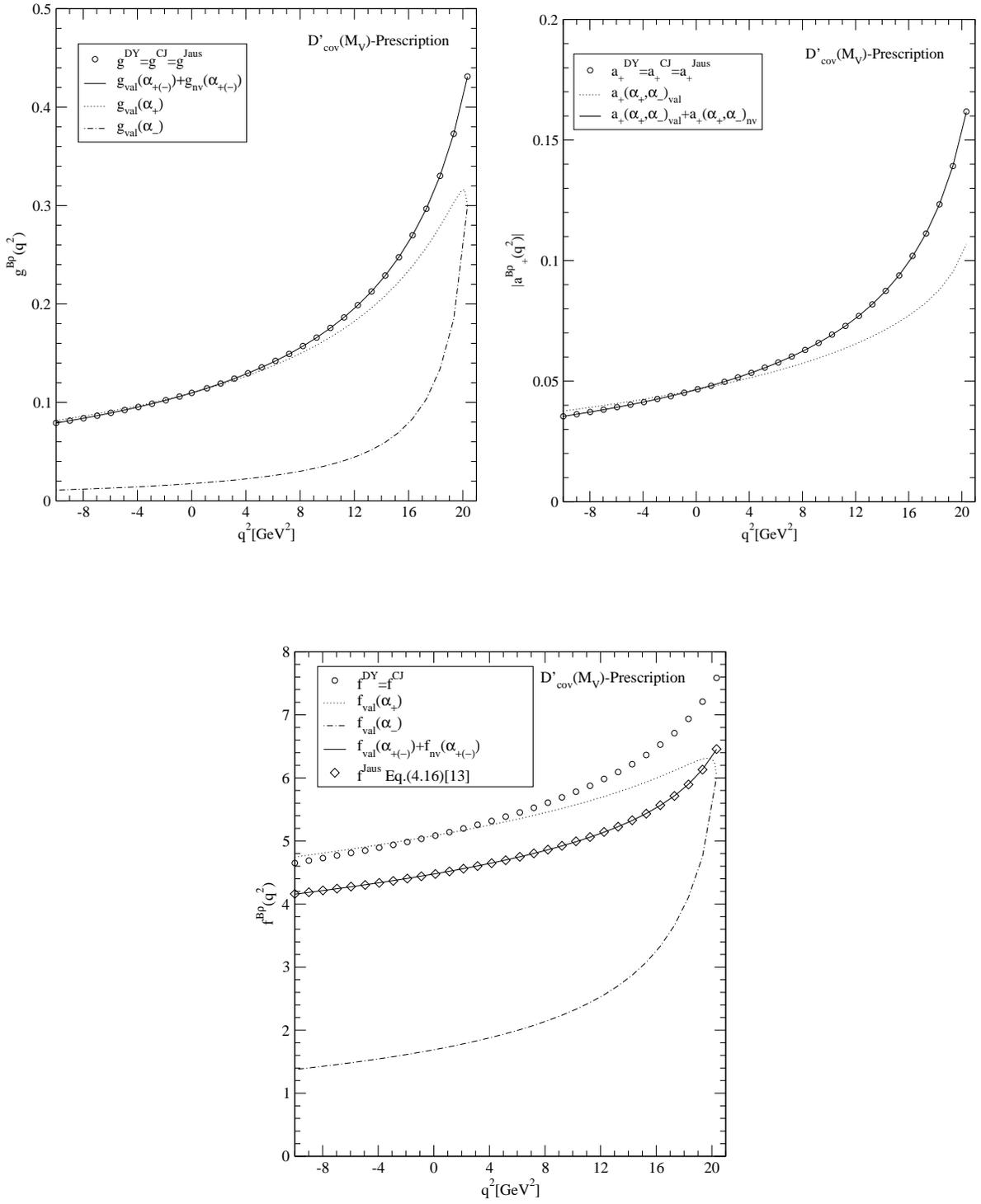

\includegraphics[width=3.0in]{gBrho_M.eps}
\hspace{0.2cm}
\includegraphics[width=3.0in]{apBrho_M.eps}
\\
\vspace{1.6cm}
\includegraphics[width=3.0in]{fBrho_M.eps}
\caption{Weak form factors $g(q^2), a_+(q^2)$ and $f(q^2)$ for 
$B\to\rho$ transition obtained from the case of the vector meson vertex
with $D_{\rm cov}(M_V)=M_V + m_2 + m$.}
\label{fig3}
\end{figure*}

In Fig.~\ref{fig3}, we present the weak form factors $g(q^2), a_+(q^2)$ 
and $f(q^2)$ for $B\to\rho$ transition obtained in the case of   
$D_{\rm cov}(M_V)=M_V+m_2 +m$,{\it i.e.} the case (1). The white circle 
represents the result in the $q^+=0$ frame obtained 
by the analytic continuation from spacelike to timelike $q^2$ region. 
We denote these form factors as $g^{\rm CJ}(q^2), a_+^{\rm CJ}(q^2)$ 
and $f^{\rm CJ}(q^2)$. For the case of $g$ form factor, we can present 
seperately $\alpha_+$ and $\alpha_-$ results for the valence contribution
since the valence calculation in the purely longitudinal frame can be done 
either by $\alpha_+$ or by $\alpha_-$ independently. 
However, this is not the case for $a_+$ form factor as shown in     
Eq.~(\ref{pl:4}). Thus, we do not separate the $\alpha_+$ result from 
$\alpha_-$ result for the form factor $a_+(q^2)$.  
For the form factor $g(q^2)$, the dotted and dot-dashed lines represent 
the valence results obtained in the purely longitudinal $q^+>0$ frame with
$\alpha_+$ and $\alpha_-$, respectively. 
The solid line represents the full(=valence + nonvalence) result
obtained from the purely longitudinal $q^+>0$ frame. 
As expected, the full result in $q^+>0$ frame is $\al_+$ and $\al_-$ 
independent.
For the form factor $a_+(q^2)$, the valence and full results are
shown by the dotted and solid lines, respectively. 
In Fig.~\ref{fig3}, we have also compared our results with the ones 
obtained from Jaus's Eqs.~(4.13),~(4.14) and~(4.16) in Ref.~\cite{Ja99} 
for the form factors $g(q^2), a_+(q^2)$, and $f(q^2)$, respectively.  
For the form factors $g(q^2)$ and $a_+(q^2)$, our results from $q^+=0$ frame, 
{\it i.e.} $g^{\rm CJ}(q^2)$ and $a^{\rm CJ}_+(q^2)$, are equivalent to Jaus's
results of $g^{\rm Jaus}(q^2)$ and $a^{\rm Jaus}_+(q^2)$ since they do not 
include $B$ and $C$-type functions in the LF integral.
Our results from $q^+=0$ frame are also in exact agreement with 
the full(valence + nonvalence) solutions obtained in the purely 
longitudinal $q^+ >0$ frame. This shows that our full results of 
$g(q^2)$ and $a_+(q^2)$ are not only Lorentz invariant but also
immune to the zero-mode contribution. 
For the form factor $f(q^2)$, however, our result $f^{\rm CJ}(q^2)$ in 
$q^+=0$ frame shows the existence of the zero-mode as explained by the 
counting rule in Sec.III.  The zero-mode contribution to 
$f^{\rm CJ}(q^2)$,{\it i.e.} the difference between the full solution(solid 
line) in $q^+>0$ frame and  $f^{\rm CJ}(q^2)$ in $q^+=0$ frame, is as large 
as 19$\%$ in the case (1).
The zero-mode contribution to $f^{DY}(q^2)$ in $q^+=0$ frame is
distinguished from that appears in the purely longitudial $q^2=q^+q^-$ frame. 
In the purely longitudinal frame, the zero-mode contribution
is a single point at $q^2=0$ as $q^+\to 0$(or equivalently $\al_+\to 1$), 
which can be quantified by the difference between 
$f_{\rm full}=f_{\rm val}(\al_+)+f_{\rm nv}(\al_+)$(solid line) 
and
$f_{\rm val}(\al_+)$(dotted line) at $q^2=0$,{\it i.e.}
$f^{\rm z.m.}(0)=\lim_{\al_+\to 1}f_{\rm nv}(\al_+)$. 
We thus distinguish the zero-mode contribution at $q^+=0$ from the usual 
nonvalence one at $q^-=0$(or equivalently $\al_-(0)$)~\cite{CJ98}.
Interestingly, however, Jaus's result(diamond) is exactly
the same as ours in $q^+>0$ frame. This indicates that his
method of including the zero-mode contribution to $f(q^2)$ 
is valid in the case (1)(or $D_{\rm cov}(M_V)$) as we have
discussed in the previous section using our power-counting method. 

\begin{figure*}
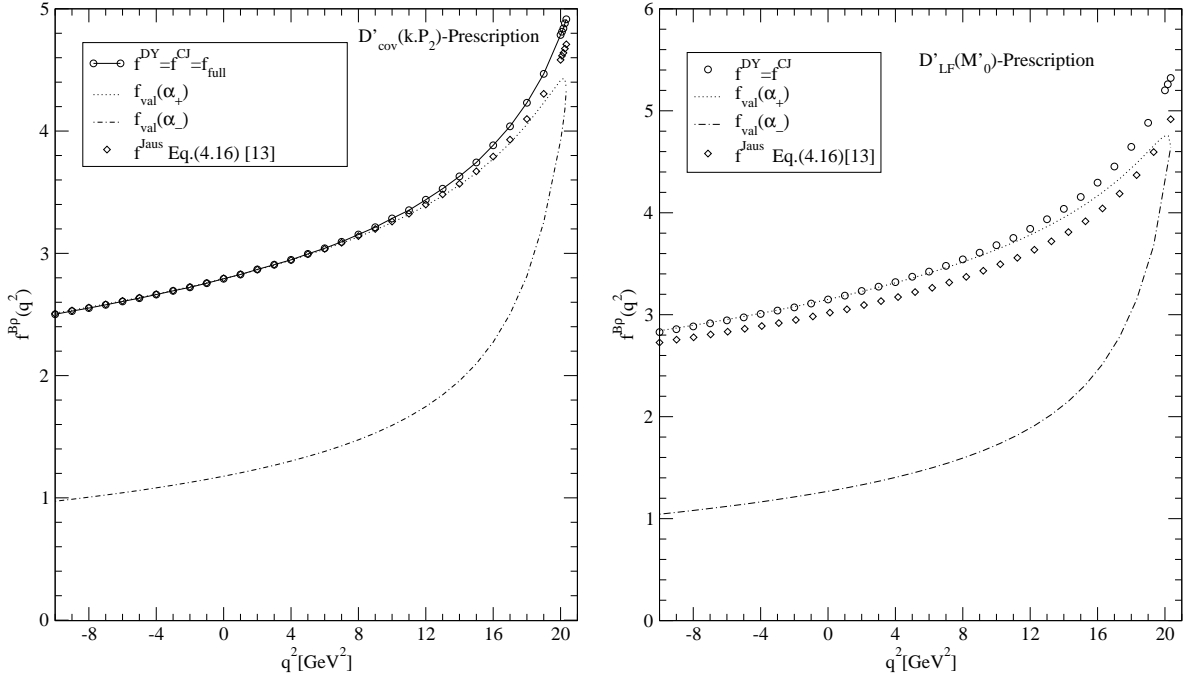

\includegraphics[width=3.0in]{fBrho_Melo.eps}
\hspace{0.2cm}
\includegraphics[width=3.0in]{fBrho_M0.eps}
\caption{Weak form factor $f(q^2)$ for $B\to\rho$ transition in the
cases of the vector meson vertex with $D_{\rm cov}(k\cdot P_2)$(left) and 
$D_{LF}(M'_0)$(right).}
\label{fig4}
\end{figure*}

In Fig.~\ref{fig4}, we present the form factor $f(q^2)$ 
for $B\to\rho$ transition in the cases (2)(or 
$D_{\rm cov}(k\cdot P_2)$(left)) and (3)(or $D_{LF}(M'_0)$(right)).
For the manifestly covariant case (2), our result 
$f^{\rm CJ}(q^2)$(circle) obtained in the $q^+=0$ frame is in an 
exact agreement 
with the full result(solid line) in the purely longitudinal $q^+>0$ frame. 
This shows that there is no zero-mode contribution to $f^{\rm CJ}(q^2)$ for
the vertex with $D_{\rm cov}(k\cdot P_2)$.  
The difference between the full result and  
$f_{\rm val}(\alpha_+)$(dotted line) or $f_{\rm val}(\alpha_-)$(dot-dashed line)
is not the zero-mode contribution but the nonvalence contribution as 
described in Fig.~\ref{fig3}.  Comparing Jaus's result with ours, 
we find that $f^{\rm CJ}(q^2)$ and $f^{\rm Jaus}(q^2)$ 
coincide each other at $q^2=0$ but differ about $4\%$ at 
$q=q_{\rm max}$. This difference between Jaus's and ours is caused by the 
different treatment of $N_2/D$ term as we have illustrated in the 
previous section(Sec.III).
Now, in the case (3) with $D_{LF}(M'_0)$, we compared the
result $f^{\rm CJ}(q^2)$(circle) in the $q^+=0$ frame with 
the valence results obtained in the purely longitudinal $q^+>0$ frame
with $\alpha_+$(dotted line) and $\alpha_-$(dot-dashed line), respectively.
As we discussed in the previous section using the power-counting rule,
the result $f^{\rm CJ}(q^2)$ without the zero-mode contribution must be 
identical to the full result. Thus, the differences between $f^{\rm CJ}(q^2)$
and the valence results(dotted line and dot-dashed line) in the $q^+>0$ 
frame exhibit the nonvalence contributions with $\alpha_+$ and $\alpha_-$,
respectively. Also, the comparison between our full result($f^{\rm CJ}$)
and Jaus's result(diamond) indicate the more substantial(even at $q^2=0$) 
difference due to the different treatment of $N_2/D$ term in the case
(3) than in the manifestly covariant case (2). For both cases of
$D_{LF}(M'_0)$ and $D_{\rm cov}(k\cdot P_2)$, we have confirmed 
that the two results(Jaus and CJ) exactly coincide if and only if we set 
$N_2/D\doteq 0$ in Jaus's formulation.

\section{Conclusion}
In this work, we have analyzed the zero-mode contribution to the weak
transition form factors, in particular $f(q^2)$, between pseudoscalar 
and vector mesons. For the phenomenologically accessible vector meson
vertex $\Gamma^\mu=\gamma^\mu - (P_2-2k)^\mu/D$, we discussed the three 
typical cases of the $D$-term which may be also classified by the 
differences in the power-counting of the LF energy $k^-$,{\it 
i.e.}: 
(1) $D_{\rm cov}(M_V)=M_V + m_2 + m\sim (k^-)^0$,
(2) $D_{\rm cov}(k\cdot P_2)=
[2k\cdot P_2 + (m_2 + m)M_V -i\epsilon]/M_V\sim (k^-)^1$,
and (3) $D_{LF}(M'_0)=M'_0 + m_2 + m\sim (k^-)^{1/2}$.
Our main idea to obtain the weak transition form factors is first to
find if the zero-mode contribution exists or not for the given form 
factor using the power-counting method. If exists, then the separation   
of the on-mass-shell propagating part from the off-mass-shell part
is useful since the off-mass-shell part is responsible for  
the zero-mode contribution. We found that the form factors $g(q^2)$ and 
$a_+(q^2)$ are immune to the zero-mode contribution in all three cases.
However, the existence/non-existence of the zero-mode in the
form factor $f(q^2)$ depends on the cases. While the zero-mode contribution
exists in the case (1) with $D_{\rm cov}(M_V)$, the other two cases
(2) and (3) with $D_{\rm cov}(k\cdot P_2)$ and $D_{LF}(M'_0)$, 
respectively, are immune to the zero-mode contribution.

This contrasts to Jaus's approach~\cite{Ja99,Ja03}. 
Although Jaus and we both agree on the vanishing zero-mode contribution to 
the form factors $g(q^2)$ and $a_+(q^2)$,
the two approaches led to different conclusions on the form
factor $f(q^2)$. 
While Jaus concluded that $f(q^2)$ receives the zero-mode contribution 
for any $D$-term, we showed that the validity of his 
prescription on $N_2/D$-term 
is limited to the case (1) with $D_{\rm cov}(M_V)$.
This is also supported by our confirmation that the two approaches 
coincide if and only if $N_2/D\doteq 0$ (not by his presciption 
$N_2/D\doteq Z_2/D$) for the cases (2) and (3) 
with $D=D_{\rm cov}(k\cdot P_2)$ and $D_{LF}(M'_0)$, respectively. 

All of these findings stem from the fact that the zero-mode 
contribution to the form factor $f(q^2)$ is absent if the denominator $D$ 
of the vector meson vertex $\Gamma^\mu=\gamma^\mu - (P_V 
-2k)^\mu/D$ contains the term proportional to the LF energy 
$(k^-)^{n}$ with the power $n>0$. Since the phenomenologically accessible 
LFQM satisfies this condition $n>0$, only the valence contribution
obtained in the $q^+ = 0$ frame is sufficient to provide the full
results of the LFQM. This certainly benefits the hadron phenomenology.

\acknowledgements
This work was supported in part by the
Korea Research Foundation(KRF-2004-002-C00063),
the grant from the U.S. Department of Energy
(DE-FG02-96ER40947), and the National Science Foundation (INT-9906384).
CRJ thanks to the hospitality provided by the Department of Physics at 
Seoul National University where he took a sabbatical leave for the spring 
semester of the year 2005 and completed this work.

\appendix
\begin{widetext}
\section{Trace terms in Eq.~(\ref{jnv})}
To obtain the nonvalence contribution to each form factor in $q^+>0$ frame,
we used the trace terms in Eq.~(\ref{jnv}) which are summarized explicitly 
in this Appendix. 

For $g_{\rm nv}(q^2)$, we need the transverse polarization with
the vector current. Thus, the trace term 
$(S^+_{h=+})_{\rm nv}(k^-_{\Lambda_1})$ for the vector current
in Eq.~(\ref{jnv}) is given by
\bea\label{A1}
[(S^+_{h=+})_{\rm nv}(k^-_{\Lambda_1})]_V &=&
-\frac{2P^+_1}{\sqrt{2}}\varepsilon^{+-xy}\biggl\{
q^L{\cal A}_p 
+ k^L [(m-m_2) + \al(m_1-m)] + \frac{2}{D}[{\bf k}^2_\perp q^L 
- ({\bf k}_\perp\cdot{\bf q}_\perp)k^L]\biggl\}.
\eea
The trace term $[(S^+_{h=+})_{\rm nv}(k^-_{m_1})]_V$ has the same form 
as the one in Eq.~(\ref{A1}).  

For $a_+(q^2)_{\rm nv}$, we need the transverse polarization with
the axial current. Thus, the trace term 
$(S^+_{h=+})_{\rm nv}(k^-_{\Lambda_1})$ for the axial current
in Eq.~(\ref{jnv}), is given by
\bea
[(S^+_{h=+})_{\rm nv}(k^-_{\Lambda_1})]_A
&=& \frac{4P^+_1}{\sqrt{2}}\biggl\{
(1-2x')q^L{\cal A}_p 
+ k^L [ (\al-2x)(m_1-m) - (m_2+m)]
-\frac{2(x'q^L + k^L)}{x'D}
\nonumber\\
&&\times\biggl({\bf k}_\perp\cdot{\bf k'}_\perp
+ [(1-x')m - x'm_2]{\cal A}_p
+x(1-x)(1-x')(M^2_1-M^2_{\Lambda_1})\biggr)
\biggr\}.
\label{A2}
\eea
The trace term $[(S^+_{h=+})_{\rm nv}(k^-_{m_1})]_A$ can be 
obtained by the replacement $\Lambda_1 \rightarrow m_1$ in Eq.~(\ref{A2}).

For $f(q^2)_{\rm nv}$, we need the longitudinal polarization with
the axial current. Thus, the trace term 
$(S^+_{h=0})_{\rm nv}(k^-_{\Lambda_1})$ for the axial current
in Eq.~(\ref{jnv}) is given by
\bea\label{A3}
[(S^+_{h=0})_{\rm nv}(k^-_{\Lambda_1})]_A
&=&-\frac{4P^+_1}{x'M_V}\biggl\{
{\cal A}_p [x'(1-x')M^2_V + m_2 m + x'^2{\bf q}^2_\perp ]
+ {\bf k}^2_\perp (xm_1 + m_2 - xm)
\nonumber\\
&&+ x'{\bf k}_\perp\cdot{\bf q}_\perp [ 2x(m_1-m) + m_2 + m ] 
+ x(1-x)m_2 (M^2_P - M^2_{\Lambda_1})
\nonumber\\
&&-\frac{1}{(1-x)D} 
\biggl( (1-x)(x'M^2_V - \alpha M^2_P) 
+ \alpha(\Lambda^2_1 + {\bf k}^2_\perp) - x'(1-x){\bf q}^2_\perp
\nonumber\\
&&- 2(1-x){\bf k}_\perp\cdot{\bf q}_\perp\biggr)
\biggl( {\bf k}_\perp\cdot{\bf k'}_\perp 
+ [ (1-x')m-x'm_2 ]{\cal A}_p + x(1-x)(1-x')(M^2_P -M^2_{\Lambda_1})
\biggr)
\biggr\},
\nonumber\\
\eea
where $M_P = M_1$ and $M_V = M_2$.
The trace term $[(S^+_{h=0})_{\rm nv}(k^-_{m_1})]_A$ can be
obtained by the replacement $\Lambda_1 \rightarrow m_1$ in Eq.~(\ref{A3}).

\end{widetext}

\end{document}